\documentclass{ws-ijmpd}

\begin{document}

\markboth{Sigismondi, Raponi, Bazin and Nugent}
{Towards a Unified Solar Limb Definition}

%
\catchline{}{}{}{}{}
%

\title{TOWARDS A UNIFIED DEFINITION OF SOLAR LIMB DURING CENTRAL ECLIPSES AND DAILY TRANSITS}

\author{Costantino Sigismondi}

\address{Sapienza University of Rome, Physics Dept. P.le Aldo Moro 5 \\
Roma, 00185, Italy\\
University of Nice-Sophia Antipolis (France); IRSOL, Istituto Ricerche Solari di Locarno (Switzerland)\\
sigismondi@icra.it}

\author{Andrea Raponi}

\address{Sapienza University of Rome, Physics Dept. P.le Aldo Moro 5 \\
Roma, 00185, Italy\\
andr.raponi@gmail.com}

\author{Cyril Bazin}

\address{Institute d´Astrophysique de Paris, 98 Bis Bd. Arago\\
Paris, F-75014, France\\
bazin.cyrille@neuf.fr}

\author{Richard Nugent}

\address{IOTA, International Occultation Timing Association, US Section\\rnugent@wt.net}

\maketitle

\begin{history}
\received{22 Mar 2011}
\revised{Day Month Year}
\comby{Managing Editor}
\end{history}

\begin{abstract}

The diameter of the Sun has been measured using Baily's beads during central eclipses, observed with portable telescopes.
A blend of tiny emission lines produced in the first several hundred kilometers above the photosphere gives a light signal which prolonges the light curves of the beads. The simple criterion of light OFF/ON adopted in the previous approaches to define the timing of photosphere's disappearance/reappearance is modified.
The technique of the limb darkening function reconstruction from the Baily's beads light curves is introduced here.
  
\end{abstract}


\section{Introduction}

The classic definition of solar limb is the location of the maximum of the derivative of the luminosity profile of the continuum spectrum,\cite{Hill75} it was adopted in the studies on solar oblateness. The application of the FFTD algorithm\cite{Hill75} allowed to avoid the seeing effects of atmospheric turbulence,\cite{Neckel95} which shift inwards the position of this maximum (identified also as the inflection point) of about $0.05$ arcsec for each arcsecond of seeing.
This value has been obtained by anti-Fourier transform (with discrete FFT with 1024 points, resolution 0.01 arcsec) of the analytic fit of the limb darkening function published by Rogerson,\cite{Rogerson} convoluted with a Gaussian with 1, 2 and 3 arcsec of FWHM: the inwards shift of the observed diameter (the distance between the two inflection points) was respectively of 0.07 arcsec, 0.18 arcsec and 0.28 arcsec.
The transition region between the photosphere and the chromosphere, under the inner corona, corresponds to the location of this inflexion point.

The location of the solar limb is different for different wavelengths due to the different height of the emitting regions: $\pm0.07$ arcsec in the domain of visible light from photosphere;\cite{Neckel94} observing the Sun in the H$\alpha$ line from above the chromosphere the solar limb is shifted by $+6.7$ arcsec corresponding to 4900 km.\cite{Wittmann73} Therefore it is necessary to identify the passing waveband with an appropriate filter in order to select the reference wavelength for measuring the solar diameter, and to avoid influence by active regions located at higher regions than the continuum spectrum coming from the solar photosphere.

\section{Problems of Limb Detections During Eclipses}
During a total eclipse the luminosity of the environment drops suddendly of several orders of magnitude as the last Baily's bead disappears. For this reason the signature of totality, especially if observed near the edges of umbral zone,\cite{Dunham73} has been considered of great utility for the measurement of the diameter, even if the observers used only naked eyes. This is the case of the eclipse of 1715 registered by Edmund Halley and discussed in a paper entitled {\it Observations of a probable change in the solar radius between 1715 and 1979}.\cite{Dunham80}
The International Occultation Timing Association (IOTA) over the last 3 decades has organized expeditions to observe the Baily's beads effect for central and annular eclipses. For the eclipses of 2005, 2006 and 2008, IOTA observers provided nearly 600 Baily's beads timings from 28 stations, with the results published in "Atlas of Baily's Beads 2005-2008".\cite{Sigismondi09}

The mismatch between some of the observing stations of the same eclipse enligthen us on the existence of a problem in the limb definition, which becomes evident when the luminosity of the photosphere is eliminated by the interposition of the Moon.

Usually the total eclipses are observed with portable equipment which does not exceed 20 cm of diameter. One of us (Sigismondi) has examined recent eclipse videos,\cite{Sigismondi09} and found that the larger the telescope aperture is, the longer the Baily's beads lasts. Here are some examples:

Joan Rovira and Carles Schnabel observed from the 2005 annular eclipse shadow limit in Spain. Here they recorded the tiny light connecting the last beads, which belongs to this intermediate region above the limb and under the chromosphere.\cite{Schnabel09} 

Let's call it the {\it marginal lines-emission region}, in continuity with the {\it marginal zones of the Moon} of the classical work of Chester B. Watts.\cite{Watts} 

Again in 2008 Chuck Herold\footnote{http://www.youtube.com/watch?v=unV18b0pKEc} observed with a 13 cm Celestron telescope the total eclipse in China, and Richard Nugent\footnote{http://www.youtube.com/watch?v=sJaNqg9WPSI} using a 9 cm Questar was 1.6 Km further out toward the shadow limit. It came out that Herold recorded the {\it marginal lines-emission region} connecting all active beads, and it remained visible also during the totality, while Nugent did not.
Since Herold was closer to the eclipse central path than Nugent, totality was longer as expected, but the begining of totality occurred without the usual zero luminosity signal.

\section{The Marginal Lines-Emission Region in the Solar Mesosphere}

The flash spectrum is obtained with a spectrograph during total eclipses, and the continuum of the spectrum corresponds to the Baily's beads (the photosphere limb with Frauhhofer lines seen in absorption), and also a myriad of faint emission lines are seen simultaneously superposed on this continuum. The thin layer associated with these faint emission lines between the photosphere (F-lines) and the low chromosphere, corresponds to the transition region which can be seen only during the internal contacts of a total eclipse. This transition region has to be taken into account while defining the solar edge, and the continuum spectrum has to be measured between these faint emission lines for accurately measuring the true solar diameter.
A moving plate spectrograph has been used in 1905 eclipse by W. W. Campbell\cite{Campbell1906} director of the Lick Observatory, in order to photograph the evolution of the spectrum with the reduction of the area of the visible Sun. 
It is visible the transition between absorption dark lines (photosphere with Fraunhofer spectrum) and tiny emission lines and chromosphere lines (few and brilliant), and finally the corona continuum. 

The blend of these tiny emission lines just above the solar limb is perceived as white light, and even if it is about 1000 times dimmer than the photosphere brightness it can be confused with the photosphere itself, when the photosphere is seen through an area 1000 times smaller and without spectrograph, as it occurs during the last stages of a total eclipse. This successfull 1905 eclipse expedition, sponsored by Crocker, remained a cornerstone for the following decades.

Nowadays the definition of reversing layer is no longer physically satisfying.
  
This region, in analogy with the situation in the Earth's atmosphere, can be defined as the {\bf solar mesosphere}.
There the temperature has a local minimum of about 4500 K instead of 5800 K of the photosphere, after the temperature rises at the beginning of the chromosphere.
This region, over the photosphere, starts with a layer where the contrast of the granules changes.
The vertical magnetic field between granules and supergranules emerges and it spreads horizontally.
The propagating waves along the magnetic field become shock waves, and they release the energy in these layers.

\begin{figure}[top]
\centerline{\psfig{file=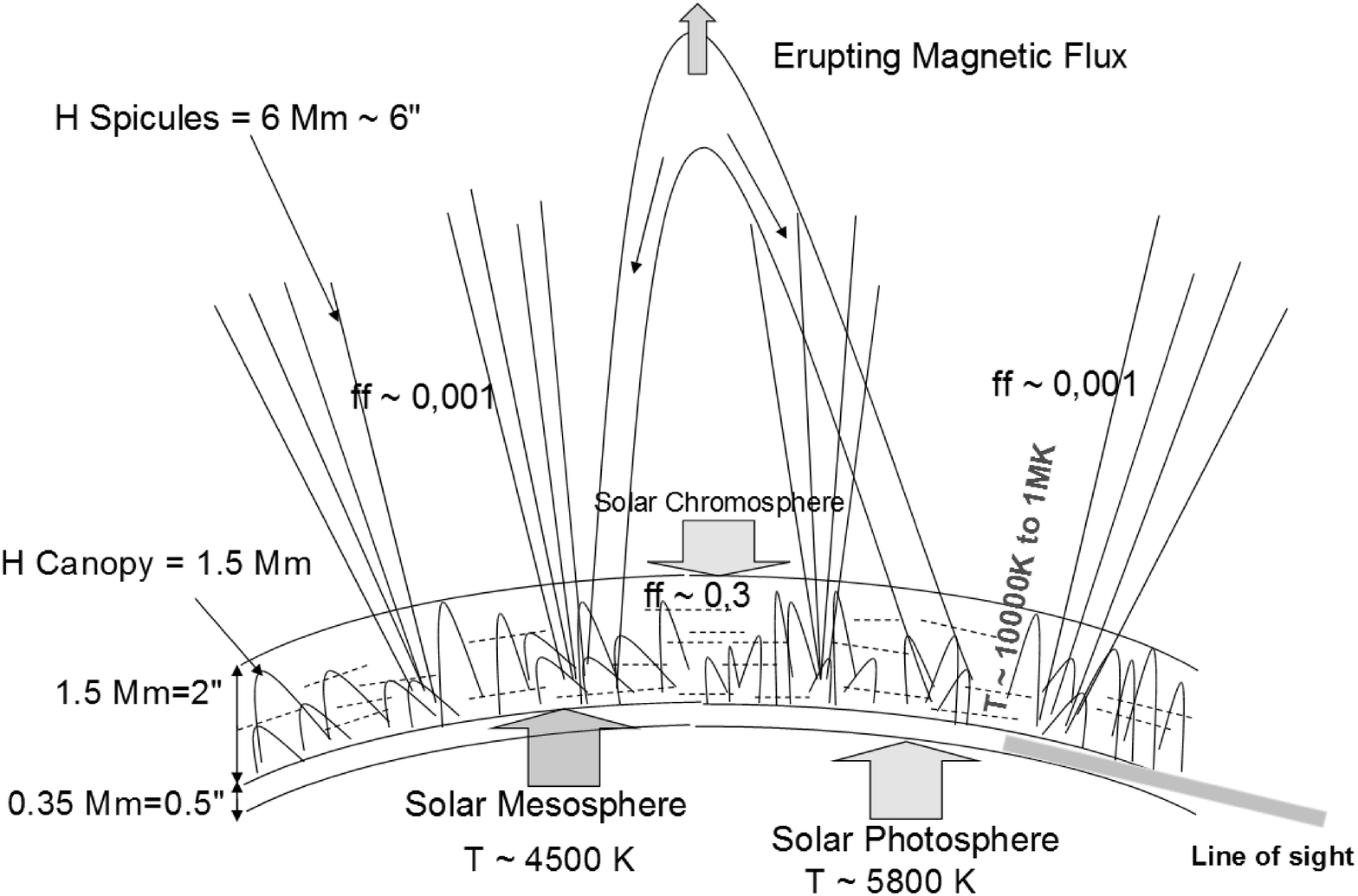,width=13cm}}
\vspace*{0pt}
\caption{The Filling Factor (FF) or heterogeneity factor, is the ratio between the volume occupied by the plasma over the total volume, because the plasma is magnetically confined. The inversion of the temperature occurs around 350 Km, or 0.50 arcsec above the photosphere.}
\end{figure}

\begin{figure}[top]
\centerline{\psfig{file=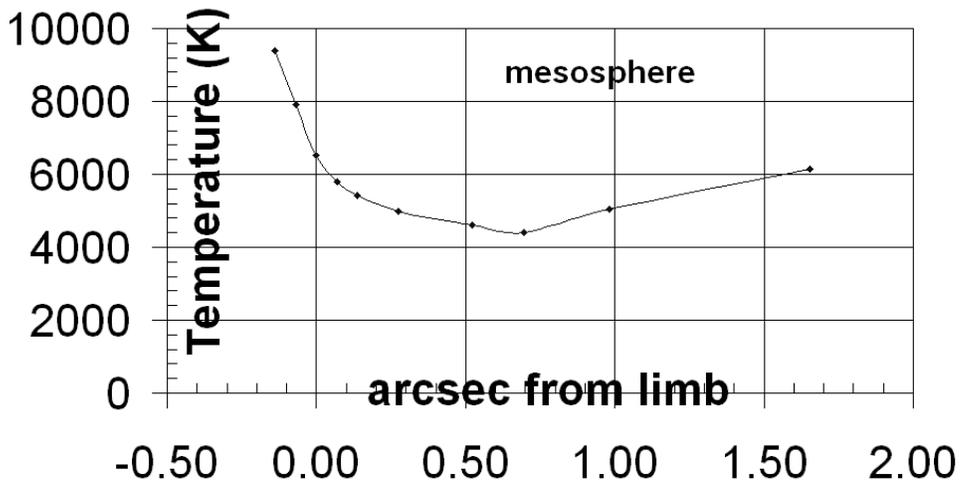,width=13cm}}
\vspace*{0pt}
\caption{In an hydrostatic one-dimensional model (data plotted from tab. 5.2 of Foukal textbook [12]) the inversion of the temperature occurs around 350 Km, or 0.50 arcsec above the photosphere.}
\end{figure}

\section{Limb Darkening Function from the Baily's Beads}

Recently, the technique of Baily's beads\cite{Dunham73} considered their signal as "ON/OFF", as if the limb darkening function would be similar to the Heaviside profile. 
This is an approximation of the real situation, otherwise an observer who is more inward in the umbral shadow should see a longer darkness during the eclipse, regardless of telescope used.
The idea, under development, is to recover the profile of the limb darkening function from the light curve of every single bead.
Frame by frame during an emerging bead event, a deeper layer of photosphere enters into the profile drawn by the lunar valleys of the limb (see Figure 3), and each layer casts light through the same geometrical area of the previous one.
The sampling is determined by the angular amplitude of the geometrical areas. The limit of this amplitude is imposed by errors of the profile of lunar valleys, which now is very accurate thanks to the Japanese lunar probe Kaguya, with its Laser Altimeter LALT experiment.\cite{Kaguya} 

\begin{figure}[top]
\centerline{\psfig{file=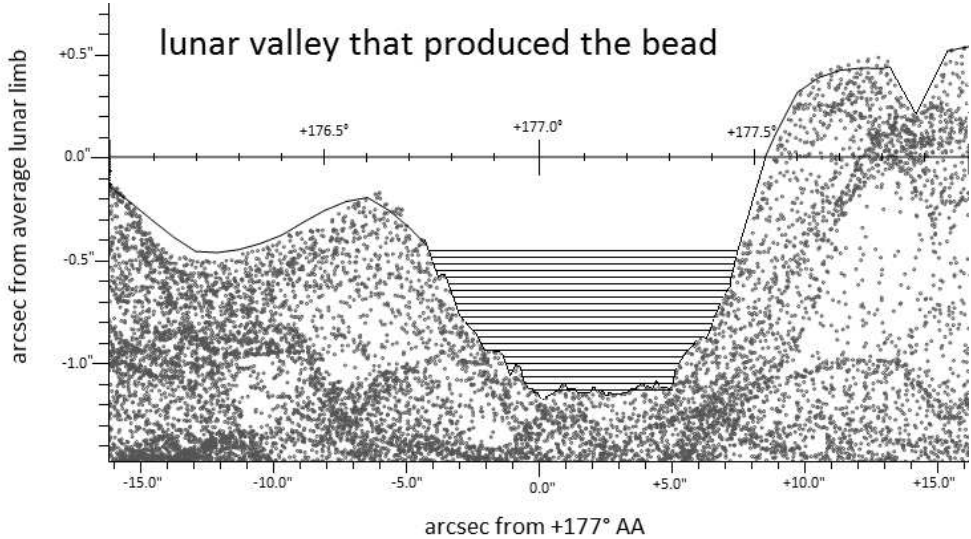,width=13cm}}
\vspace*{0pt}
\caption{The Kaguya profile of the valley at the axis angle AA=177$^\circ$ during the annular eclipse of January 15, 2010: the bead here considered 
was formed through this valley. The area of the valley is divided into 21 slices of
equal height (30 milliarcsec). During the bead's event each slice is filled with different
layers of the solar limb every time interval (1.1 s for this bead).}
\end{figure}

The sources of error for the computed luminosity function at the limb arise from the error of the Kaguya profile and from the level of signal-to-noise ratio of the light recorded from the beads, this light curve is shown in Figure 4.

\begin{figure}
\centerline{\psfig{file=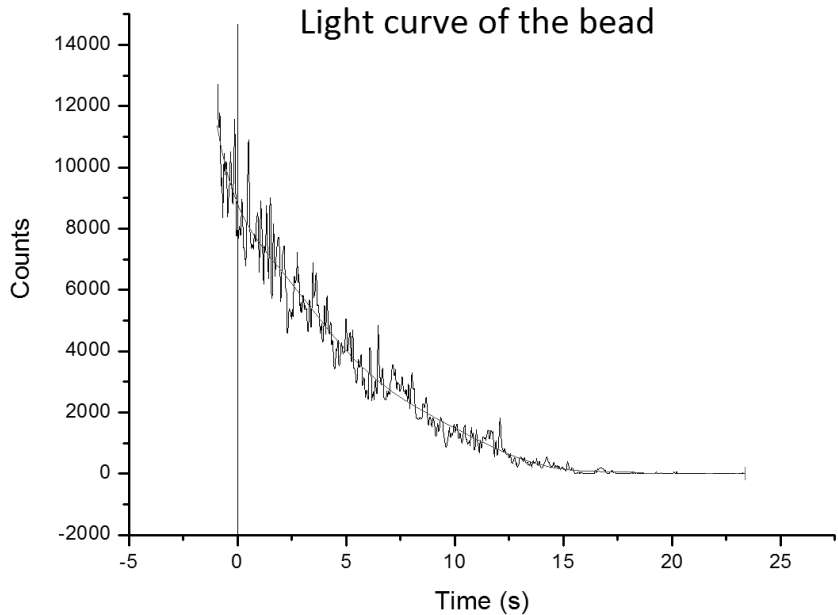,width=13cm}}
\vspace*{0pt}
\caption{The light curve of the examined disappearing bead at the axis angle AA=177$^\circ$ during the annular eclipse of January 15, 2010 is shown, with its
polynomial fit. In abscissa are the time (s), by setting to
zero the instant when all the pixels are unsaturated. The velocity of acquisition is 30 frames per second. The counts on the y axis are the sum of the intensities of all pixels involved in the bead. Each pixel has 256 levels of intensity (8 bits).}

\end{figure}

An example of the results of the procedure of computation of the limb darkening function from a Baily's bead is shown in the Figure 5.

\begin{figure}
\centerline{\psfig{file=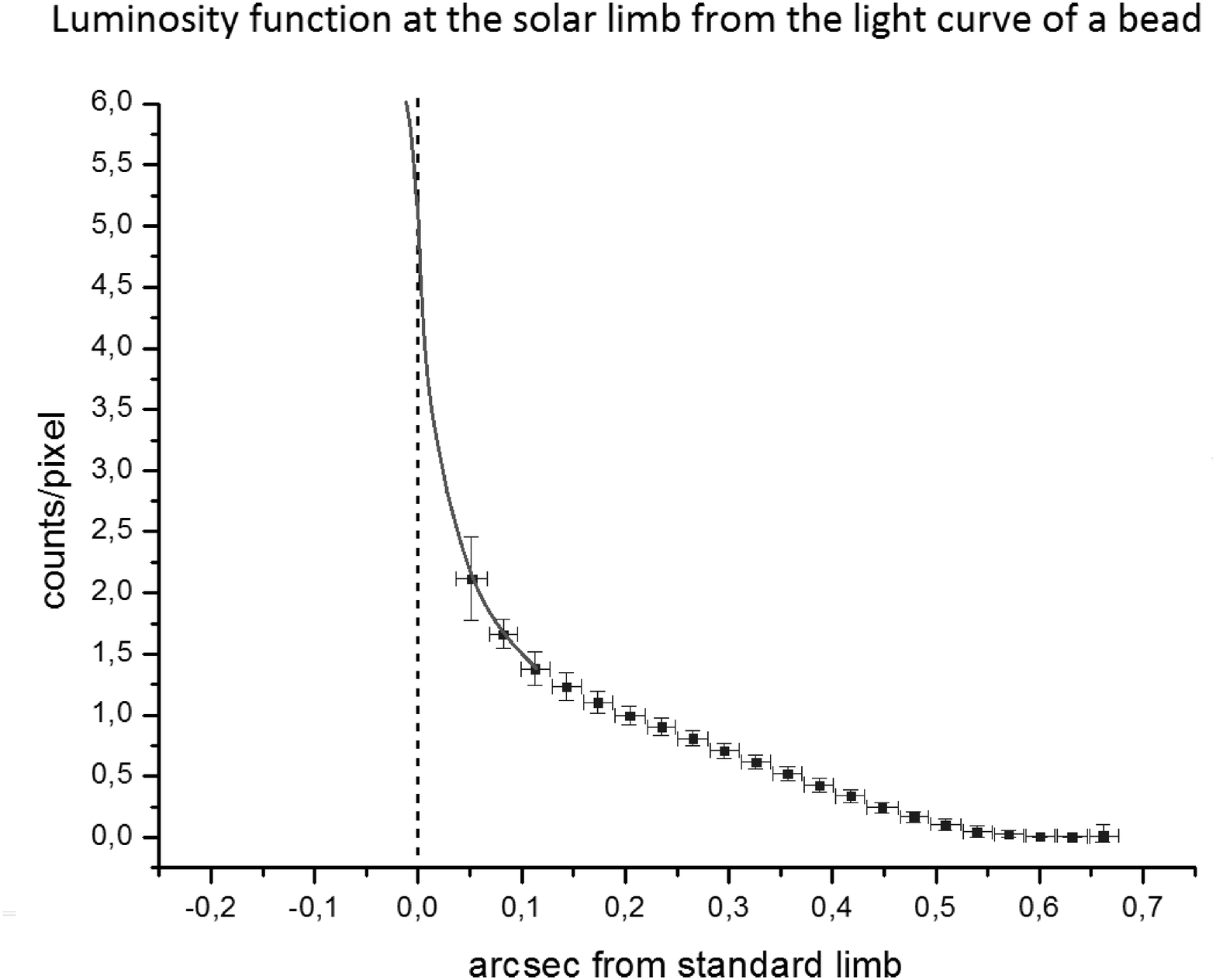,width=13cm}}
\vspace*{0pt}
\caption{Luminosity function at the solar limb from the light curve of a bead. The result of the analysis of the light curve of a disappearing Baily's bead observed during the annular eclipse made on January 15, 2010 in
Uganda by Richard Nugent is shown. The zero of the abscissa is the position of the standard solar limb with
a radius of 959.63 arcsec at 1 AU. The y axis is the solar surface
brightness in term of CCD counts divided by the involved area of the lunar
valley where the bead is formed.
The saturation of the CCD avoided to measure the luminosity function more
inwards, and the solid line gives a possible scenario on the position of the
inflection point.}
\end{figure}

The form of the limb darkening function found shows that almost all of the unsatured light curve of this bead is external to the inflection point i.e. the solar limb. The standard solar limb is located at the radius equal to 959.63 arcsec.\cite{auwers}
This value has been obtained after the application of an "irradiation correction" of 1.55 arcsec to the visual observations. This correction is due to the diffusion into the atmosphere, because the point spread function of the telescopes are different for each telescope.
Finally in Figure 6 the quantum efficiency of the detector
the Watec 902H Ultimate\footnote{http://www.aegis-elec.com/products/watec-902H\_spec\_eng.pdf} as a function of the wavelength is shown. The sensitivity (minimum illumination) of the camera is 0.3 mLux at maximum gain.

\begin{figure}
\centerline{\psfig{file=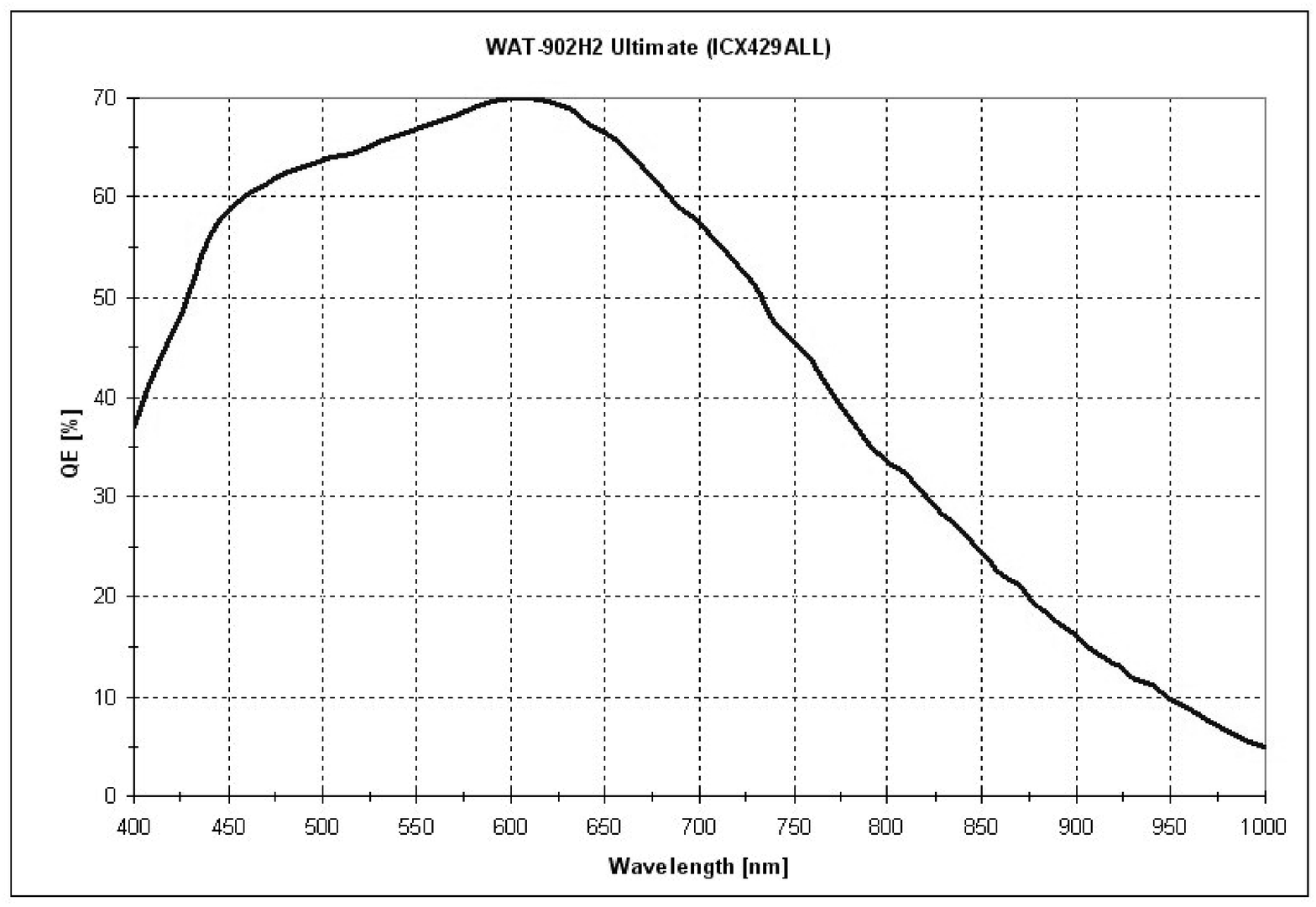,width=13cm}}
\vspace*{0pt}
\caption{Quantum efficiency as a function of the wavelength for the CCD camera Watec 902H Ultimate used during the annular eclipse of January 15, 2010.
The other filter in front of the objective of the 9 cm Questar Matsukov-Cassegrain telescope (f=1300 mm) was panchromatic Thousand Oaks density 5 with a transmittance of 1 part over $10^5$. The correction to the final position of the inflection point can be larger than the the estimated error of sampling, of $\pm0.015$ arcsec, when this quantum efficiency curve is combined with the solar spectrum.}

\end{figure}

For the Jan 15, 2010 eclipse observation made by Richard Nugent, the light curve of a disappearing bead was examined along with the valley that produced it. The resulting
points show a shape of the solar limb clearly external with respect to the inflection point. 
Therefore from these points it is impossible to infer an exact location of the inflection point, but it is possible to deduce an upper limit for it, corresponding to the innermost measured point of the curve. In terms of distance from the standard solar limb this limit was $+0.050$ arcsec $\pm0.015$ arcsec. The error is due to the sampling of the points every $0.030$ arcsec.

\section{Conclusions}
We obtained a limb darkening function with a spatial resolution of 15 milliarcsec using a standard video during an annular eclipse.
The solar limb can be placed at the inflection point of this luminosity function, using the same method adopted for the observations of the full disk of the Sun in the classical works on solar oblateness or in the measurements of the solar diameter with drift-scan methods.\cite{clavius08} 
The unified definition of solar limb is related to the identification of the inflection point, or the maximum of the first derivative, of the luminosity profile of the Sun. This criterion becomes identical in the eclipses and in the daily transits, once the limb darkening function is computed by the Baily's beads light curves.
In some sense the luminosity profile obtained during the eclipse is the limit for zero seeing of the daily profile, since the geometry of the Sun-Moon is determined outside of the atmosphere and because of the elimination of all scattered light by the atmosphere, which is screened by the Moon.
The method of computation of the limb darkening function using Baily's beads is very promising, especially if we consider that, in terms of the signal-to-noise ratio, total eclipses are better than the annular ones. An increased sensitivity of the CCD or CMOS  detectors from 8 bits corresponding to 256 levels to 12 bits, which correspond to 4096 level of detectable intensity, is recommended. In this way we can extend the sampling of the luminosity function to regions of the photosphere more internal and luminous than the inflection point, without being saturated before it. 
 
\section*{Acknowledgments}

Many thanks to Alberto Egidi (Roma, Tor Vergata), Serge Koutchmy (Institute of Astrophysique de Paris), Michele Bianda and Renzo Ramelli (IRSOL, Locarno), David Dunham (IOTA), Konrad Guhl (IOTA and Berlin Archenhold Sternwarte) and Carles Schnabel (IOTA and Sabadell Astronomical Association) for the fruitful collaboration on this particular topic.

\end{document}